\documentclass{article}
\usepackage{spconf,amsmath,graphicx}

\usepackage{booktabs}
\usepackage{multirow}
\usepackage{array}
\usepackage{siunitx}
\usepackage{threeparttable}
\title{Scaling Audio Models Efficiently: Joint Optimization of Scale, Resolution, Adaptation, Precision, and Sparsity}
\name{Vyom Agarwal$^{1}$, Mokshda Gangrade$^{1}$, Siddharth Pal$^{3}$, Jerry Wu$^{2}$}
\address{
  $^{1}$Computer, Mathematical, and Natural Sciences, University of Maryland, College Park, MD, USA\\
  $^{2}$Electrical and Computer Engineering, University of Maryland, College Park, MD, USA\\
  $^{3}$RTX BBN Technologies, Columbia, MD, USA\\
  vyomwal5@umd.edu, mokshdag@umd.edu, siddharth.pal@rtx.com, jerrywu@umd.edu
}

\begin{document}
\ninept
\maketitle
\begin{abstract}
Large automatic speech recognition (ASR) models such as Whisper must be deployed across hardware with widely varying memory and inference-speed constraints. We present a compression framework that jointly parametrizes Whisper deployment along \emph{six} dimensions: model size, temporal resolution, encoder token stride, low-rank adaptation capacity, weight precision and sparsity pattern. All axes are jointly optimized using NSGA-III with respect to three deployment objectives: word error rate (WER), inference FLOPs, and memory footprint. Across 50 of the 1,680 candidate configurations evaluated, we characterize the conditional effect of each axis and identify compression combinations that dominate naive single-axis scaling, while finding that 1:4 structured sparsity fails to recover acceptable accuracy under the tested recovery budgets. We report measured WER and resident memory, use analytical EffFLOPs as the search-time compute surrogate, and separately validate representative inference configurations using measured real-time factor (RTF).
\end{abstract}
\section{Introduction}
\label{sec:intro}

Whisper and other large ASR models achieve strong accuracy during transcription but require compute costs and memory that make direct deployment impractical on edge hardware or real time inference constraints. A large body of work addresses individual axes of this problem: low rank adaptation~\cite{hu2022lora} reduces finetuning cost without affecting inference cost, post-training and quantization-aware training reduce weight precision, structured N:M sparsity reduces the number of non-zero weights while remaining amenable to hardware acceleration on modern GPUs. USM-Lite~\cite{ding2024usmlite} combines quantization and sparsity in a single joint finetuning recipe and is the closest prior work to this paper. However, USM-Lite and the broader literature typically study these axes two at a time and do not jointly consider architectural axes like Model Size, temporal resolution and encoder token stride alongside precision and sparsity in a single search space.

We argue that a practitioner implementing a Whisper deployment configuration is not choosing a quantization level in isolation: rather, they are simultaneously deciding on a model size, context length, encoder stride, adapter capacity, precision and sparsity and these choices interact with one another and collectively motivate the contributions of this paper:

\begin{enumerate}
\item We define and evaluate a \textbf{six-axis compression space} (Sec.~\ref{sec:expmethod}) spanning model size ($x_N$), input length ($x_T$), encoder token stride ($x_V$), adapter rank ($x_R$), weight precision ($x_Q$) and sparsity pattern ($x_P$), and report the effect of each axis on three deployment objectives independently (Sec.~\ref{sec:peraxis}).
\item We frame configuration selection as \textbf{three-objective Pareto optimization} (word error rate, FLOPs, memory), rather than a single scalarized cost, using NSGA-III~\cite{deb2014nsga3} with LibriSpeech \emph{test-clean} WER as the search objective and \emph{test-other} for final evaluation (Sec.~\ref{subsec:method}).
\item We report \textbf{measured deployment accuracy and memory} from genuinely converted inference artifacts and complement the analytical EffFLOPs objective with representative RTF measurements on an NVIDIA RTX A6000, including an optimized CTranslate2 INT8 deployment (Sec.~\ref{sec:deployment}).
\end{enumerate}
\begin{figure}[t]
  \centering
  \includegraphics[width=1\columnwidth]{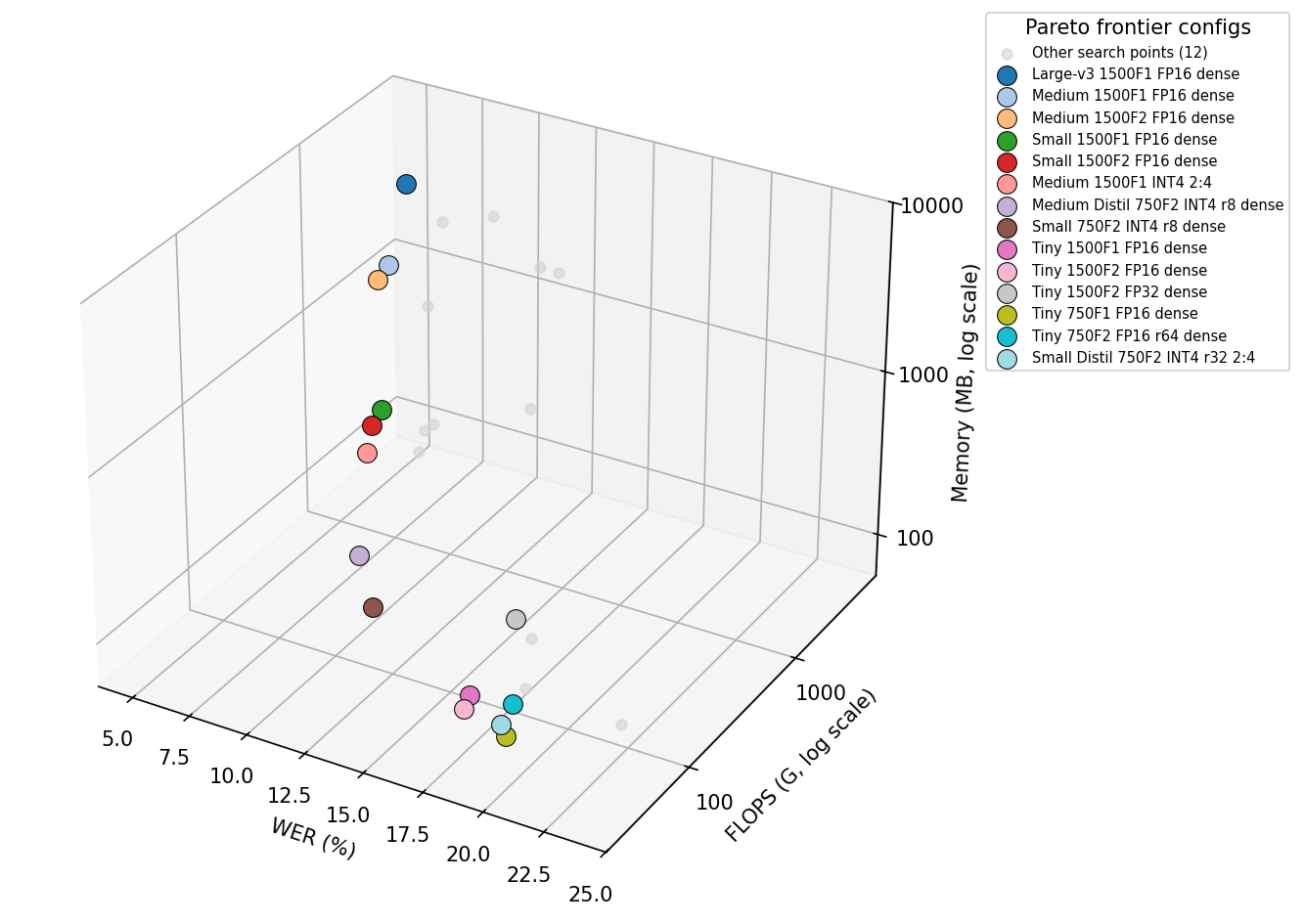}
  \caption{Three-objective Pareto front over all evaluated
  configurations. Non-dominated points (highlighted) trace the joint
  WER/FLOPs/memory trade-off space. Background points are the dominated configurations.}
  \label{fig:3D}
\end{figure}

\section{Related Work}
\label{sec:related}

Whisper~\cite{radford2023whisper} is an encoder-decoder Transformer
trained on large-scale weakly-supervised speech data, released at several model sizes (tiny through large-v3). LoRA~\cite{hu2022lora} enables parameter efficient adaptation by injecting trainable low-rank updates into frozen attention projections. Structured N:M sparsity, in which $N$ of every $M$ contiguous weights are pruned, is directly supported by Ampere-generation sparse Tensor Cores for the 2:4 pattern~\cite{nvidia2020a100}, giving a real inference speedup on compatible hardware. USM-Lite~\cite{ding2024usmlite} jointly finetunes a 2-billion parameter universal speech model under quantization and N:M sparsity constraints, reporting that a single compression technique alone fails to recover performance under extreme settings (int2 quantization, 1:4 sparsity), while their joint approach compresses to 9.4\% of original size at a 7.3\% relative WER cost. Our sparsity findings (Sec.~\ref{sec:peraxis}) extend this observation to a different model family, a broader range of model sizes, both adapter based and full parameter recovery regimes.
Weight only 4-bit quantization via the NF4 data type~\cite{dettmers2023qlora} is designed around the empirical observation that trained neural network weights are approximately normally distributed and is the mechanism we use for genuinely deployed INT4 evaluation. Concurrent work has explored finer grained, per-layer mixed-precision
quantization learned jointly with training for speech foundation models~\cite{xu2025mixedprecision}. Our study instead evaluates a small, fixed set of precisions post-hoc across a broader six dimension deployment space.

\section{Methodology}
\label{sec:expmethod}
We parameterize a Whisper deployment configuration along six dimensions,
summarized in Table~\ref{tab:axes}.

\begin{table}[t]
\centering
\caption{The six-axis compression space.}
\label{tab:axes}
\resizebox{\columnwidth}{!}{%
\begin{tabular}{@{}llp{4.7cm}@{}}
\toprule
\textbf{Axis} & \textbf{Values} \\
\midrule
Model Size ($x_N$) & tiny, small, medium, large-v3, distil-small, distil-medium, distil-large \\
Input length ($x_T$) & 750 / 1500 encoder frames truncating input audio to 15 s / 30 s \\
Encoder token stride ($x_V$) & 1 / 2 (post encoder subsampling of decoder cross attention length) \\
Adapter rank ($x_R$) & LoRA rank $r \in \{0(\text{no lora}),8,16,32,64\}$ \\
Weight precision ($x_Q$) & FP32, FP16, INT8, INT4 \\
Sparsity pattern ($x_P$) & dense, 2:4, 1:4 (N:M structured) \\
\bottomrule
    \end{tabular}}
\end{table}
The full grid spans 1,680 candidate configurations; we evaluate 50 out of them via the NSGA-III guided search described in Sec.~\ref{subsec:method}. Each axis is motivated by a distinct deployment constraint: $x_N$ trades raw capacity for footprint, $x_T$ trades acoustic context for encoder compute, $x_V$ trades decoder cross attention length for decoder compute without touching encoder cost, $x_R$ trades adaptation capacity for a marginal parameter and memory overhead, $x_Q$ and $x_P$ directly trade numeric and structural redundancy for memory and compute.

Training follows a two-stage process, a dense (or precision aware) finetuning stage on LibriSpeech \texttt{train.100},
followed by a magnitude(1:4/2:4) based one-shot pruning step and a shorter recovery finetuning stage under a frozen sparsity mask ($x_P \neq$ dense). Pruning is deferred to after convergence rather than applied at initialization, following standard practice for stable N:M sparsity recovery.

\subsection{Whisper Model}
\label{subsec:whispermodel}

Whisper~\cite{radford2023whisper} is an encoder-decoder Transformer
for speech recognition and translation. Input audio is resampled to
16\,kHz and converted to an 80-channel log-Mel spectrogram over a
fixed 30-second window (3000 frames at a 10\,ms hop), which two 1D
convolutional layers project into the encoder's input dimension, the
second using stride~2 and halving the sequence length to 1500
positions before a stack of standard Transformer self-attention
encoder blocks. The decoder is an autoregressive Transformer that
attends to the encoder output via cross-attention and predicts a
byte-pair token sequence. We evaluate the four multilingual
checkpoint sizes (tiny, small, medium, large-v3), spanning 39M to
1.5B parameters, as well as their English-only distilled
counterparts (distil-small, distil-medium, distil-large), which
reduce decoder depth while keeping the encoder intact. Throughout
this paper, $x_T$ (Sec.~\ref{sec:peraxis}) refers to the number of
\emph{post-convolution encoder positions} obtained by truncating the
input audio itself before feature extraction (e.g., 15\,s of audio
yields 750 post-convolution positions rather than the standard
1500). $x_V$
(Sec.~\ref{sec:peraxis}) operates on this same post-convolution
sequence, subsampling it before it reaches decoder cross-attention.

\subsection{Three Objective Optimization Framework}
\label{subsec:method}

We treat configuration selection as multi-objective optimization over
three deployment objectives, computed per configuration -

\begin{itemize}
\item \textbf{Word Error Rate (WER)}~\cite{morris2004wer}, evaluated on LibriSpeech \emph{test-clean}~\cite{panayotov2015librispeech} during NSGA-III search and used as the search-time accuracy objective. LibriSpeech \emph{test-other} is withheld from the search procedure and used only for final evaluation of the selected configurations.
\item \textbf{Effective inference FLOPs (EffFLOPs)}, an analytical
compute-cost surrogate. We first compute architectural FLOPs, reduce
operations associated with pruned linear layers according to their
retained weight ratio, and apply the nominal precision and sparsity
factors in Table~\ref{tab:scales}. These factors represent idealized
arithmetic scaling rather than measured FLOPs. We additionally benchmark
representative configurations on an NVIDIA RTX A6000 where FP32/FP16 and an
optimized CTranslate2 INT8 deployment are measured directly, while the
INT4 value remains an idealized projection outside the scope of this study.
\item \textbf{Resident weight memory (MB)}, measured directly from the resident parameter and buffer tensors of the \emph{actual} model object used at inference time. This measures static weight storage only. Critically, for INT8 and INT4 configurations measured on a genuinely converted artifact (Sec.~\ref{sec:deployment}). For brevity, we write ``memory'' or ``Mem'' as shorthand for this quantity throughout the remainder of the paper.
\end{itemize}
\begin{table}[t]
\centering
\caption{Nominal factors used by the analytical EffFLOPs surrogate and
representative RTX A6000 RTF. ``M'' denotes measured runtime and ``I''
an idealized projection. INT8 uses an optimized CTranslate2 backend.}
\label{tab:scales}
\scriptsize
\setlength{\tabcolsep}{3pt}
\renewcommand{\arraystretch}{1.0}
\begin{tabular}{@{}llccc@{}}
\toprule
Axis & Setting & Eff. scale & RTF \\
\midrule
\multirow{4}{*}{$x_Q$}
& FP32 & $1.0\times$   & 0.0534 \\
& FP16 & $0.5\times$   & 0.0454 \\
& INT8 & $0.25\times$  & 0.0263 \\
& INT4 & $0.125\times$ & 0.0113(Ideal) \\
\midrule
\multirow{3}{*}{$x_P$}
& Dense & $1.0\times$  & 0.0454 \\
& 2:4   & $0.5\times$  & 0.0272 \\
& 1:4   & $0.25\times$ & -- \\
\bottomrule
\end{tabular}
\end{table}

We use NSGA-III~\cite{deb2014nsga3}, a reference-point-based
non-dominated sorting genetic algorithm for optimizing $\geq3$
objectives, seeded with all configurations evaluated to date and
re-run after each batch to propose the next. The 50 configurations
were selected across 5 successive batches of roughly 10. The
reported front reflects a single NSGA-III trajectory; a new
trajectory requires fresh training runs, so sensitivity to search
initialization remains a limitation of this study.

\section{Six Dimensions and their Analysis}
\label{sec:peraxis}

We first isolate the conditional effect of each axis, holding the other five axes at a fixed default reference configuration, before considering joint interactions in Sec.~\ref{sec:deployment}. The default parameters considered are Model Size, 1500 frames, 1 token stride, finetune with LoRA r32, FP16 and no pruning. WER values reported in the following analysis are measured on LibriSpeech \emph{test-other}.

\textbf{Model Size ($x_N$)}:
Table~\ref{tab:xn} confirms the expected monotonic tradeoff: WER, FLOPs and memory all scale together with model capacity. The FLOPs range spans 60$\times$ (73.1\,G to 4393.8\,G) for a 3.5$\times$ relative WER improvement (16.48\% to 4.85\%), illustrating that model size alone is a comparatively inefficient axis for the last few points of WER improvement relative to the compute it costs.

\begin{table}[t]
\centering
\caption{Model Size $x_N$ (1500 frames, r32, stride 1, dense, FP16).}
\label{tab:xn}
\resizebox{\columnwidth}{!}{%
\begin{tabular}{@{}lrrr@{}}
\toprule
Model & WER (\%) & Eff. FLOPs (G) & Mem (MB) \\
\midrule
Tiny (39M)     & 16.48 & 73.06   & 75.7  \\
Small (244M)   & 7.96  & 645.45  & 474.6 \\
Medium (769M)  & 5.61  & 2161.32 & 1492.9 \\
Large-v3 (1.5B) & 4.85 & 4393.76 & 3010.4 \\
\bottomrule
\end{tabular}}
\end{table}

\textbf{Input length ($x_T$)}:
Halving the input audio duration (30s to 15s, correspondingly 1500 to 750 post-convolution encoder positions) reduces FLOPs by about 37-44\% across model sizes, as expected from the encoder's quadratic self-attention dependence on $x_T$
(Table~\ref{tab:xt}). The WER cost of this reduction grows with model size in relative terms (15.4\% for Tiny vs.\ 54--60\% for Medium and Large-v3), though the absolute WER increase is comparatively consistent across scales (2.5-3 percentage points). Our interpretation is that \emph{larger models rely more heavily on extended acoustic context to resolve ambiguity smaller models cannot exploit regardless.}

\begin{table}[t]
\centering
\caption{Input length $x_T$ (stride 1, dense, FP16). $\Delta$WER
is relative to the 1500-frame configuration.}
\label{tab:xt}
\resizebox{\columnwidth}{!}{%
\begin{tabular}{@{}lrrrr@{}}
\toprule
Model & $x_T{=}750$ & Eff. FLOPs $\Delta$ & WER $\Delta$ (rel.) \\
\midrule
Tiny     & 19.01\% & $-36.9\%$ & $+15.4\%$ \\
Small    & 10.84\% & $-43.0\%$ & $+36.2\%$ \\
Medium   & 8.64\%  & $-43.8\%$ & $+54.0\%$ \\
Large-v3 & 7.76\%  & $-43.7\%$ & $+60.0\%$ \\
\bottomrule
\end{tabular}}
\end{table}

\begin{figure}[t]
  \includegraphics[width=0.8\columnwidth]{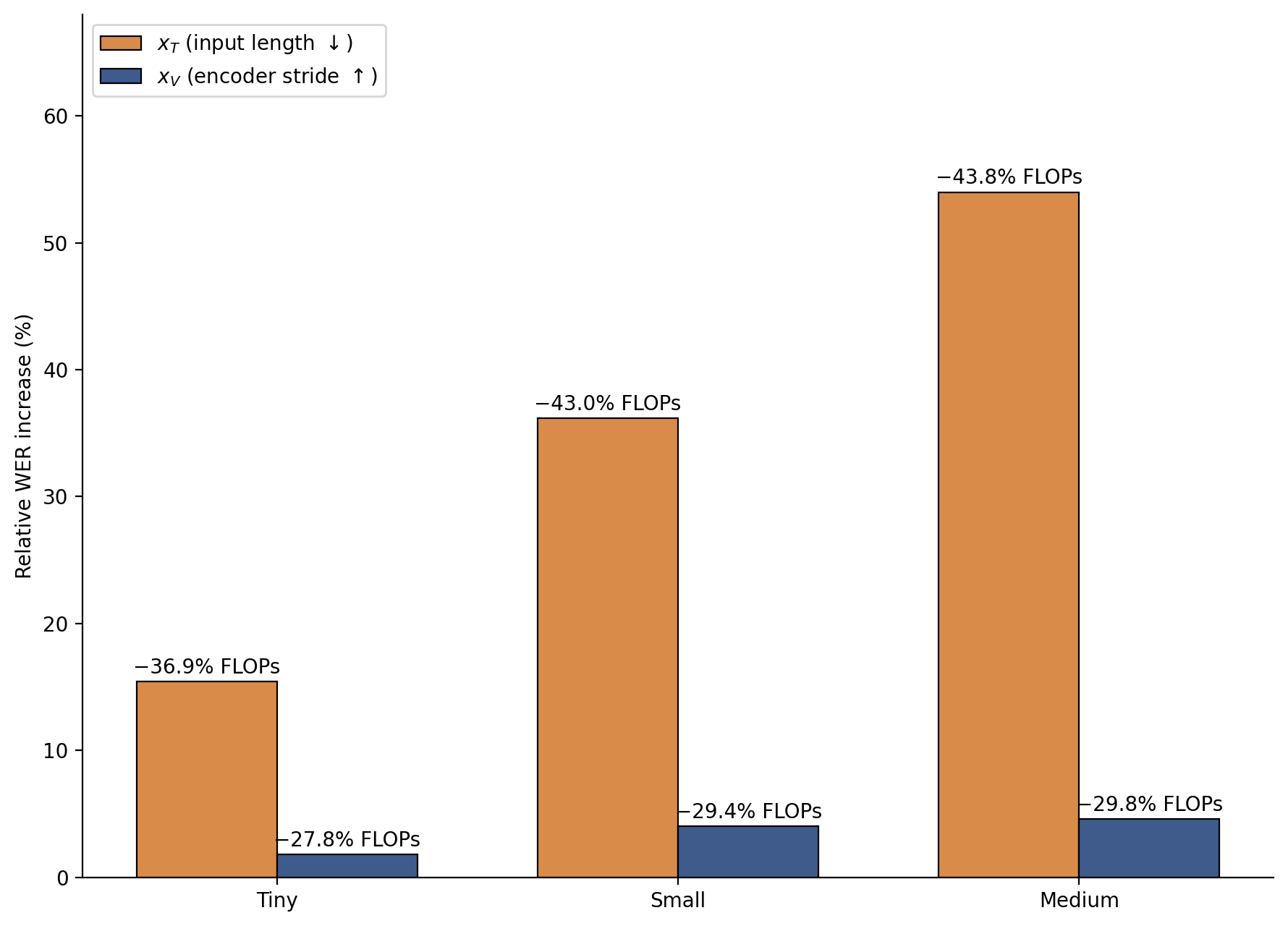}
  \caption{Relative WER increase vs.\ FLOPs reduction for encoder
  token stride ($x_V$) and input length ($x_T$) at matched model
  scales. For a comparable FLOPs cut, $x_V$ costs a fraction of the
  WER that $x_T$ does at every scale tested.}
  \label{fig:xvxt}
\end{figure}

\textbf{Encoder token stride ($x_V$)}:
Fig.~\ref{fig:xvxt} shows a markedly more favorable tradeoff than $x_T$: subsampling the encoder output by a factor of 2 before it reaches decoder cross-attention reduces the total FLOPs by 22–29\% at every model size tested, at a relative WER cost under 5\% in every case. This makes the parameter more favorable than reducing input duration. This asymmetry follows directly from where each axis acts (Sec.~\ref{subsec:whispermodel}), leaving the encoder's own representation capacity untouched.

\textit{Practical implication:} for a fixed FLOPs budget, $x_V$ should be exhausted as a compression lever before $x_T$ is reduced, a
configuration selection not evident when either axis is studied in isolation, but only visible once both are varied against the same WER objective jointly.

\textbf{Adapter rank ($x_R$)}: 
Adapter rank targets only the query and value attention projections in
this study, so its direct memory overhead is small relative to base model
size at every rank tested. Its primary observed role in this study is being the mechanism by which a compressed (pruned and/or quantized) model recovers accuracy when the base weights themselves are frozen. In matched configurations holding all other five axes fixed, increasing rank from $r{=}16$ to $r{=}32$ (Tiny, INT4/1:4) and $r{=}32$ to $r{=}64$ (Large-v3, INT4/1:4) did not improve recovery, indicating adapter capacity alone cannot compensate for the representational damage 1:4 sparsity causes at attention key/output and feed-forward projections it leaves untouched.

\textbf{Weight precision ($x_Q$)}:
Table~\ref{tab:xq} reports two representative deployed
configurations (each jointly varying $x_T$/$x_V$/$x_P$ alongside
precision). INT8 on Medium gives a $1.66\times$ memory reduction relative to FP16, while INT4 (NF4) on Small with 2:4 sparsity gives $2.67\times$. These fall below the nominal $2\times$/ $4\times$ reductions because only linear-layer weights are quantized. For NF4, the 4-bit factor reflects storage precision, not necessarily native 4-bit arithmetic.

\begin{table}[t]
\centering
\caption{Measured effects for two representative deployed configurations ($x_T$/$x_V$/$x_P$ also vary). Resident memory depends only on precision; ratio is relative to dense FP16 of the same model size.}
\label{tab:xq}
\resizebox{\columnwidth}{!}{%
\begin{tabular}{@{}llrrc@{}}
\toprule
Config & Precision & WER (\%) & Mem (MB) & Mem ratio \\
\midrule
Med/1500-1/dense  & FP16 (ref.) & 5.61  & 1492.9 & 1.00$\times$ \\
Med/750-2/2:4     & real INT8   & 11.99 & 897.9  & 0.60$\times$ \\
Small/1500-1/dense& FP16 (ref.) & 7.96  & 474.6  & 1.00$\times$ \\
Small/1500-1/2:4  & real INT4   & 18.58 & 177.6  & 0.37$\times$ \\
\bottomrule
\end{tabular}}
\end{table}

One anomaly is worth reporting rather than smoothing over: Tiny under dense real INT8 quantization reached only 52.6\% WER, a substantial regression from its FP16 baseline that is not observed at any other model size under the same precision. We infer this arises due to optimistically heavy compression on a tiny model and report it as an open limitation (Sec.~\ref{sec:limitations}) to study further.

\textbf{Sparsity pattern ($x_P$)}: For this we compare deployed configurations and realize this is the sharpest asymmetry in the entire six dimension space. 2:4 structured sparsity recovers gracefully and consistently in experiments when combined with real INT8 quantization, Medium reaches 11.99\% WER (Table~\ref{tab:xq}), combined with real INT4, Small reaches 18.58\%. Both are graceful degradations from their respective dense baselines, not catastrophic failures.

1:4 sparsity, however, fails to recover acceptable accuracy in every configuration tested in this study. This holds across four model scales and all three distilled checkpoints, both adapter-based and full-parameter recovery, and every precision setting, regardless of recovery step budget up to 4000 steps. In a dedicated ablation extending the recovery budget for a
full-parameter, precision-unconstrained Medium configuration to match the \emph{original} dense pre-pruning finetuning budget (4000 steps, an 8$\times$ increase over the standard 500-step recovery window used otherwise in this study), WER still failed to recover. Training loss improved monotonically throughout this extended run, while WER simultaneously \emph{increases} after an early minimum, evidence against both explanations at the budgets we tested, though the following paragraph notes we cannot rule out recovery at substantially larger budgets.

\textbf{A note on training scale.} USM-Lite's joint compression recipe trains for 200{,}000 steps at batch size 2{,}048~\cite{ding2024usmlite} roughly $3{,}200\times$ more training exposure than our most extended $x_P$ ablation, and $25{,}600\times$ more than the standard 500 steps recovery budget used otherwise in this study. We therefore do not claim 1:4 sparsity is fundamentally unrecoverable, only that it fails within a compute budget several orders of magnitude smaller than prior work. For a deployment oriented study, that is the practically relevant finding, post-hoc compression is only useful if it recovers within the modest budget and actually spending for adapting a released checkpoint. Whether 1:4 sparsity would eventually recover at USM-Lite's scale remains open, within any realistic post-hoc compression budget, it is not currently a usable axis for Whisper.

\begin{table}[t]
\centering
\caption{Deployment recommendations with measured WER/memory and
analytical EffFLOPs trade-offs.}
\label{tab:deploy}
\footnotesize
\setlength{\tabcolsep}{3pt}
\renewcommand{\arraystretch}{1.15}
\begin{tabular}{@{}l l S[table-format=2.2] S[table-format=4.1] S[table-format=4.1]@{}}
\toprule
Target & Config & {WER (\%)} & {Mem (MB)} & {Eff. FLOPs (G)} \\
\midrule
Cloud   & L-v3/1500-1/r32/D/FP16 & 4.85  & 3010.4 & 4393.76 \\
Server  & Med/1500-1/r0/2:4/INT4    & 9.02  & 448.9  & 297.48  \\
Edge    & Small/750-2/r8/D/INT4  & 12.48 & 177.6  & 69.413   \\
Minimal & Tiny/1500-2/r32/D/FP16     & 16.77 & 75.7 & 57.069  \\
\bottomrule
\end{tabular}
\begin{tablenotes}
\footnotesize
\item L-v3 = Large-v3; Med = Medium; D = dense; r$N$ = LoRA rank $N$;
\end{tablenotes}
\end{table}
\begin{figure}[t]
  \centering
  \includegraphics[width=1\columnwidth]{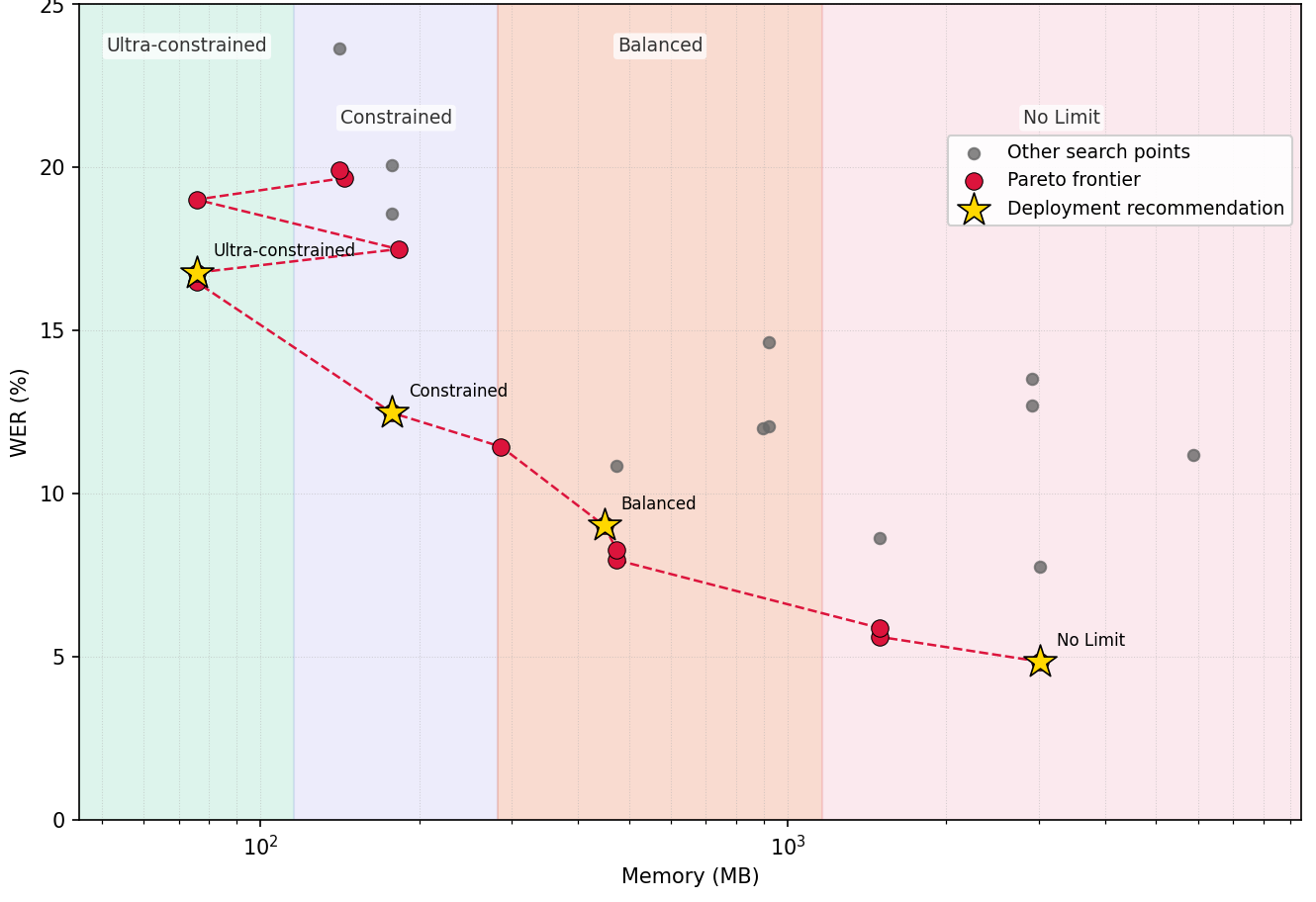}
  \caption{Pareto front in WER-vs-Memory space, annotated with the
  four deployment recommendations of Table~\ref{tab:deploy} (Cloud,
  Server, Edge, Minimal), spanning memory.}
  \label{fig:deploy2d}
\end{figure}

\section{Joint Optimization and Deployment Recommendations}
\label{sec:deployment}

Following completion of the NSGA-III search using \emph{test-clean} WER, we evaluate the selected configurations on the held-out LibriSpeech \emph{test-other} split. Table~\ref{tab:deploy} translates the resulting final WER/memory/compute trade-off region into four representative deployment targets, each with measured accuracy, memory, and effective FLOPs figures a practitioner can use
directly to select a configuration for a given hardware budget, without needing to rerun the search themselves. The choice of four deployment bands spanning cloud to severely memory-constrained targets follows the same on-device memory and power considerations that motivate architecture-level redesign in recent Transformer-based streaming ASR work~\cite{li2024folding}

Two points are worth highlighting from this table. First, the Server to Edge transition trades a further 2.5$\times$ memory reduction and 4.3$\times$ FLOPs reduction for a 3.46-point absolute WER increase, whether this is an acceptable trade depends entirely on the target application's error tolerance, which our framework deliberately leaves as a downstream decision. Second, the Minimal ultra-constrained row uses \emph{dense} FP16 Tiny rather than a quantized Tiny variant due to the INT8-on-Tiny Sec.~\ref{sec:peraxis}, a concrete illustration of \emph{why per-axis assumptions (``quantization always shrinks the model'') must be checked jointly against the other axes rather than applied uniformly.} The breadth of the Pareto region has a practical consequence: a practitioner who cannot deploy a listed configuration exactly, is likely to find a near-equivalent substitute at comparable cost.

\textit{Note on measurement.}
WER and resident-memory values are measured from the actual evaluated artifacts; EffFLOPs is analytical. We separately benchmark representative inference configurations using RTF. In particular, optimized CTranslate2 INT8 inference reaches an RTF of approximately 0.0262 on the RTX A6000, compared with 0.0454 for our FP16 reference. INT4 runtime remains an idealized projection because an optimized native INT4 backend was not implemented in this study. All memory figures above are measured
on converted artifacts, real INT8 via dynamic quantization and real
INT4 via NF4 ~\cite{dettmers2023qlora}.

\section{Conclusion}
We presented a six-dimension, three-objective compression framework for Whisper ASR deployment, characterizing the marginal effects of model size, input length, encoder stride, adapter rank, precision, and sparsity independently before jointly optimizing them through multi-objective Pareto optimization. Our central findings are 1) encoder token stride is a substantially more efficient compression axis than input length reduction at every model size tested. 2) Real, measured quantization achieves meaningfully sub-theoretical but still practically useful
compression ratios. 3) 2:4 structured sparsity composes gracefully with both LoRA adaptation and real weight quantization while 1:4 sparsity does not recover under any tested recovery strategy, corroborating prior findings on extreme single-axis compression. 4) These results translate directly into concrete, measured deployment configurations spanning cloud to edge devices with constrained targets.

\section{Limitations}
\label{sec:limitations}

The INT8-on-Tiny WER regression remains
unexplained within this study. The checkpoint trained and converged
normally, however the regression appears specifically at inference time under
real INT8 conversion, not during training. We have verified this
result across repeated independent conversions and realized we
cannot rule out that comparable, subtler effects are present elsewhere
in our INT8 results without having been large enough to notice. Our INT4 evaluations use NF4, a nonlinear quantization scheme. Finally, our 1:4 sparsity ablations, while extensive
across model size, adaptation mode, and one substantially extended recovery budget, remain bounded by the compute available to this study. We, therefore, cannot rule out that a sufficiently long joint quantization-and-sparsity recipe, matching USM-Lite's original training scale, would eventually recover 1:4 performance, only that no configuration tested here does.

\section{Acknowledgment}
\label{sec:ack}
This work was supported by compute resources provided by the University of
Maryland. No external funding was received for conducting this study, and
the authors have no relevant financial or non-financial interests to
disclose. This document does not contain technology or technical data controlled under either the U.S. International Traffic in Arms Regulations or the U.S. Export Administration Regulations.

\section{Compliance with Ethical Standards}
\label{sec:ethics}
This is a computational study using publicly available datasets
(LibriSpeech~\cite{panayotov2015librispeech}) and publicly released
pretrained model checkpoints (Whisper~\cite{radford2023whisper}). No human
or animal subjects were involved and no ethical approval was required.

\bibliographystyle{IEEEbib}
\bibliography{refs}

\clearpage

\section{Extended Ablations and Implementation Details}
\label{sec:extended_ablation}

This section expands the implementation and ablation evidence behind the six-axis search.  The purpose is to separate three quantities that are intentionally different in the main paper: \emph{EffFLOPs} is an analytical search-time compute surrogate; \emph{RTF} is a measured runtime quantity on a particular software/hardware backend; and \emph{resident weight memory} is measured static parameter/buffer storage.  Consequently, a configuration can have low analytical EffFLOPs without achieving proportionally lower wall-clock RTF when the deployed kernel does not realize the assumed low-precision or structured-sparse arithmetic.

\subsection{EffFLOPs implementation}
\label{subsec:effflops_impl}

For every configuration we first compute a dense-equivalent arithmetic count from the resolved Whisper architecture.  Let $d$ be the model width, $T$ the post-convolution encoder length, $S=\lfloor T/x_V\rfloor$ the sequence length seen by decoder cross-attention, $N_E$/$N_D$ the encoder/decoder layer counts, $d_{ff}^{E}$/$d_{ff}^{D}$ the corresponding feed-forward widths, $V$ the vocabulary size, $B$ the beam width, $P$ the decoder prompt length, and $L$ the fixed number of generated tokens.  The standardized calculation uses $B=16$, $P=1$, and $L=50$.  With KV caching, the number of decoder token positions that pass through projection and feed-forward layers is
\begin{equation}
N=P+\max(L-1,0),
\end{equation}
and the total number of self-attention query--key pairs is
\begin{equation}
C_{\mathrm{self}}=P^2+(L-1)P+\frac{(L-1)L}{2}.
\end{equation}
The convolutional stem is counted as
\begin{equation}
F_{\mathrm{conv}}=2(2T)d(80)(3)+2Td^2(3).
\end{equation}
Per encoder layer, the linear projections and feed-forward network cost
\begin{equation}
F_{E,\mathrm{lin}}=8Td^2+4Td d_{ff}^{E},
\end{equation}
while attention matrix products cost $4T^2d$.  The complete encoder contribution is therefore
\begin{equation}
F_E=F_{\mathrm{conv}}+N_E\left(8Td^2+4Td d_{ff}^{E}+4T^2d\right).
\end{equation}
The decoder calculation is KV-cache-aware.  Self-attention projections cost $8BNd^2$ per layer and self-attention matrix products cost $4BC_{\mathrm{self}}d$.  Cross-attention query/output projections cost $4BNd^2$, cached encoder key/value projections cost $4BSd^2$, and cross-attention matrix products cost $4BNSd$.  The feed-forward contribution is $4BNd d_{ff}^{D}$, and the final vocabulary projection is $2BNdV$.  Thus $x_V$ changes $S$ and therefore decoder cross-attention cost, but does not reduce the encoder computation itself.

We partition the resulting raw arithmetic into $F_{\mathrm{fixed}}$ and the base-Linear component $F_{\mathrm{lin}}$ eligible for N:M pruning.  For an ideal N:M density $\rho_P$,
\begin{equation}
F_{\mathrm{sparse}}=F_{\mathrm{fixed}}+\rho_PF_{\mathrm{lin}}+F_{\mathrm{LoRA}},
\end{equation}
with $\rho_P\in\{1,0.5,0.25\}$ for dense, 2:4, and 1:4.  $F_{\mathrm{LoRA}}$ is included separately when a sparse base must remain unmerged and the LoRA $q\_\mathrm{proj}/v\_\mathrm{proj}$ branches execute densely.  The final surrogate is
\begin{equation}
\boxed{\mathrm{EffFLOPs}=q_Q F_{\mathrm{sparse}}},
\end{equation}
where $q_Q=1,0.5,0.25,0.125$ for FP32, FP16, INT8, and INT4 respectively.  These precision factors are \emph{nominal compute-cost factors}; they do not mean that the GPU executed that fraction of the mathematical operations.  In particular, the NF4 implementation below stores 4-bit weights but performs its matrix multiplication path with FP16 compute, and the present project does not contain a native 1:4 sparse kernel.

\begin{figure*}[t]
\centering
\includegraphics[width=0.78\textwidth]{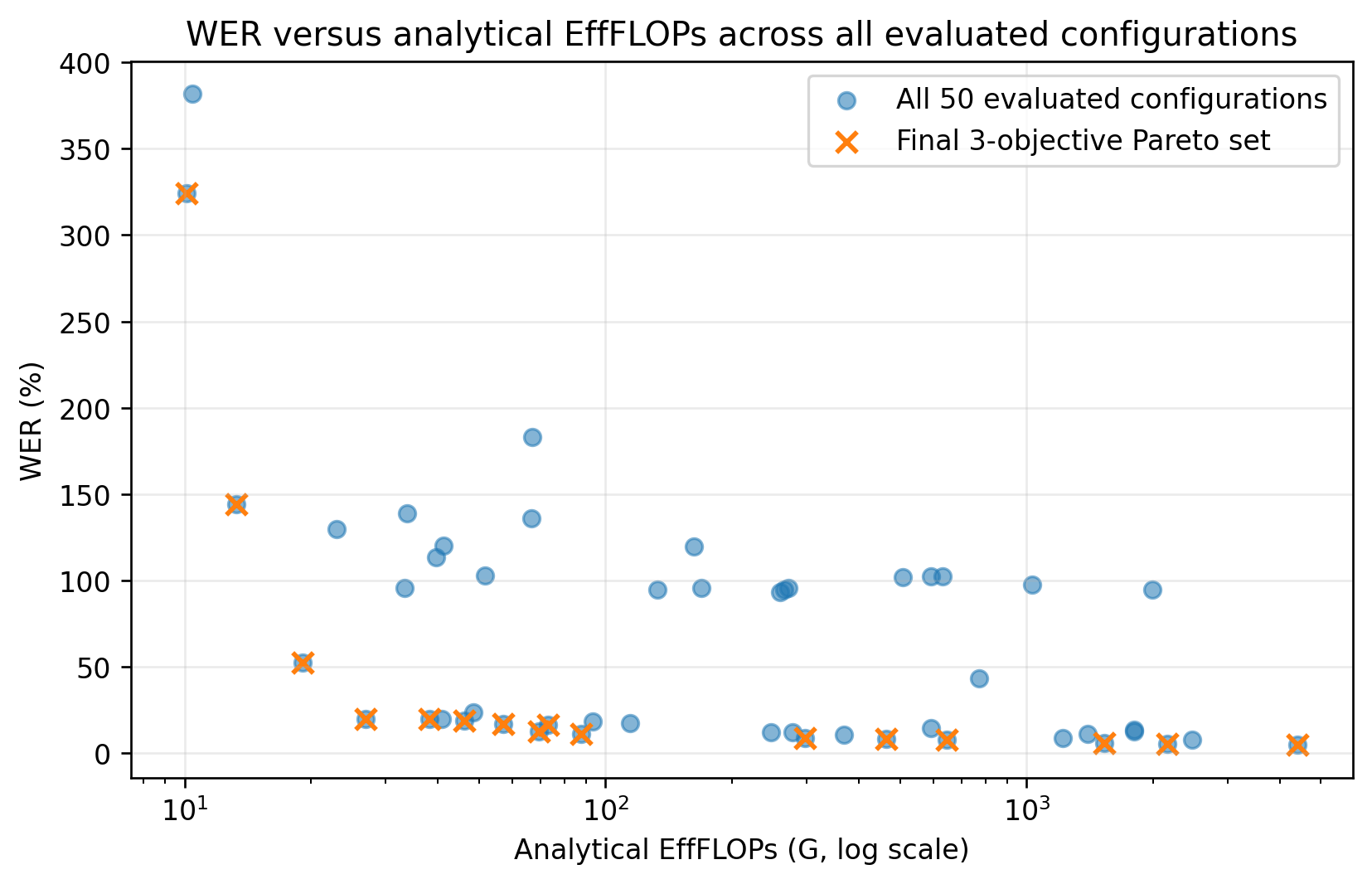}
\caption{WER versus corrected analytical EffFLOPs for all 50 evaluated configurations.  Crosses denote the final three-objective non-dominated set under WER, EffFLOPs, and resident weight memory.  The logarithmic compute axis makes clear that extreme low-compute 1:4 configurations can remain mathematically non-dominated despite unusable WER.}
\label{fig:ab_wer_eff}
\end{figure*}

\subsection{Full Pareto set and interpretation}
\label{subsec:pareto_full}

Table~\ref{tab:pareto_full} lists the final non-dominated set after recomputing all 50 configurations with the corrected EffFLOPs values.  The front is deliberately shown without imposing an application-specific WER ceiling.  Therefore some very high-WER points survive because no other evaluated configuration is simultaneously lower in WER, EffFLOPs, and memory.  These points are useful for understanding the geometry of the search but are not deployment recommendations; the main paper's Cloud/Server/Edge/Minimal table imposes practical judgment after the Pareto calculation.

\begin{table*}[t]
\centering
\caption{Final corrected three-objective Pareto set.  Precision is the $x_Q$ search label; runtime backend details are separated in Sec.~\ref{subsec:quant_impl}.}
\label{tab:pareto_full}
\scriptsize
\setlength{\tabcolsep}{3.2pt}
\begin{tabular}{@{}llrrr@{}}
\toprule
ID & Configuration & WER (\%) & EffFLOPs (G) & Resident mem. (MB)\\
\midrule
B1-1 & Tiny/750-1/r32/dense/FP16 & 19.01 & 46.107 & 75.69 \\
B1-2 & Tiny/1500-1/r32/dense/FP16 & 16.48 & 73.068 & 75.69 \\
B1-3 & Tiny/1500-2/r32/dense/FP16 & 16.77 & 57.069 & 75.69 \\
B1-5 & Small/1500-1/r32/dense/FP16 & 7.96 & 645.452 & 474.57 \\
B1-6 & Small/1500-2/r32/dense/FP16 & 8.26 & 464.524 & 474.57 \\
B1-8 & Medium/1500-1/r32/dense/FP16 & 5.61 & 2161.322 & 1492.94 \\
B1-9 & Medium/1500-2/r32/dense/FP16 & 5.88 & 1527.851 & 1492.94 \\
B1-11 & Large-v3/1500-1/r32/dense/FP16 & 4.85 & 4393.757 & 3010.39 \\
B3-2 & Medium-Distil/750-2/r8/dense/INT4 & 11.43 & 87.331 & 286.20 \\
B3-7 & Small/750-2/r8/dense/INT4 & 12.48 & 69.413 & 177.57 \\
B3-8 & Tiny/750-1/r0/1:4/INT8 & 144.08 & 13.277 & 96.80 \\
B3-9 & Tiny/750-2/r0/dense/INT8 & 52.59 & 19.054 & 96.80 \\
B4-3 & Tiny/1500-1/r16/1:4/INT4 & 324.19 & 10.091 & 48.40 \\
B4-8 & Medium/1500-1/r0/2:4/INT4 & 9.02 & 297.480 & 448.94 \\
B5-4 & Tiny/750-2/r64/dense/FP16 & 19.67 & 38.108 & 144.05 \\
B5-5 & Small-Distil/750-2/r32/2:4/INT4 & 19.93 & 26.890 & 141.37 \\
\bottomrule
\end{tabular}
\end{table*}

\subsection{Search trajectory and stopping evidence}
\label{subsec:search_stop}

The 50 real training/evaluation jobs arrived in five successive batches containing 11, 10, 10, 10, and 9 configurations.  Using the corrected objective values, the cumulative non-dominated set contains 9, 10, 14, 15, and 16 points after 11, 21, 31, 41, and 50 evaluations.  To obtain a scale-independent post-hoc convergence diagnostic, each of the three objectives was min--max normalized over the final 50 points and a fixed reference point $(1.05,1.05,1.05)$ was used to compute three-dimensional dominated hypervolume.  Relative to the final cycle, the hypervolume values are 99.119\%, 99.333\%, 99.859\%, 99.975\%, and 100\%.  Because the initial batch intentionally spans Tiny through Large-v3, it already covers most of the global objective range; later batches primarily densify the front.  The fifth cycle changes normalized hypervolume by only 0.025 percentage points while increasing the front size from 15 to 16.  This plot is a \emph{post-hoc} stopping diagnostic, not a claim that the final hypervolume was known during the search.

\begin{table}[t]
\centering
\caption{Corrected cumulative search diagnostics.}
\label{tab:search_cycles}
\scriptsize
\begin{tabular}{@{}rrrr@{}}
\toprule
Cycle & Cumulative evals & Pareto size & HV / final (\%)\\
\midrule
1 & 11 & 9  & 99.119\\
2 & 21 & 10 & 99.333\\
3 & 31 & 14 & 99.859\\
4 & 41 & 15 & 99.975\\
5 & 50 & 16 & 100.000\\
\bottomrule
\end{tabular}
\end{table}

\begin{figure}[t]
\centering
\includegraphics[width=\columnwidth]{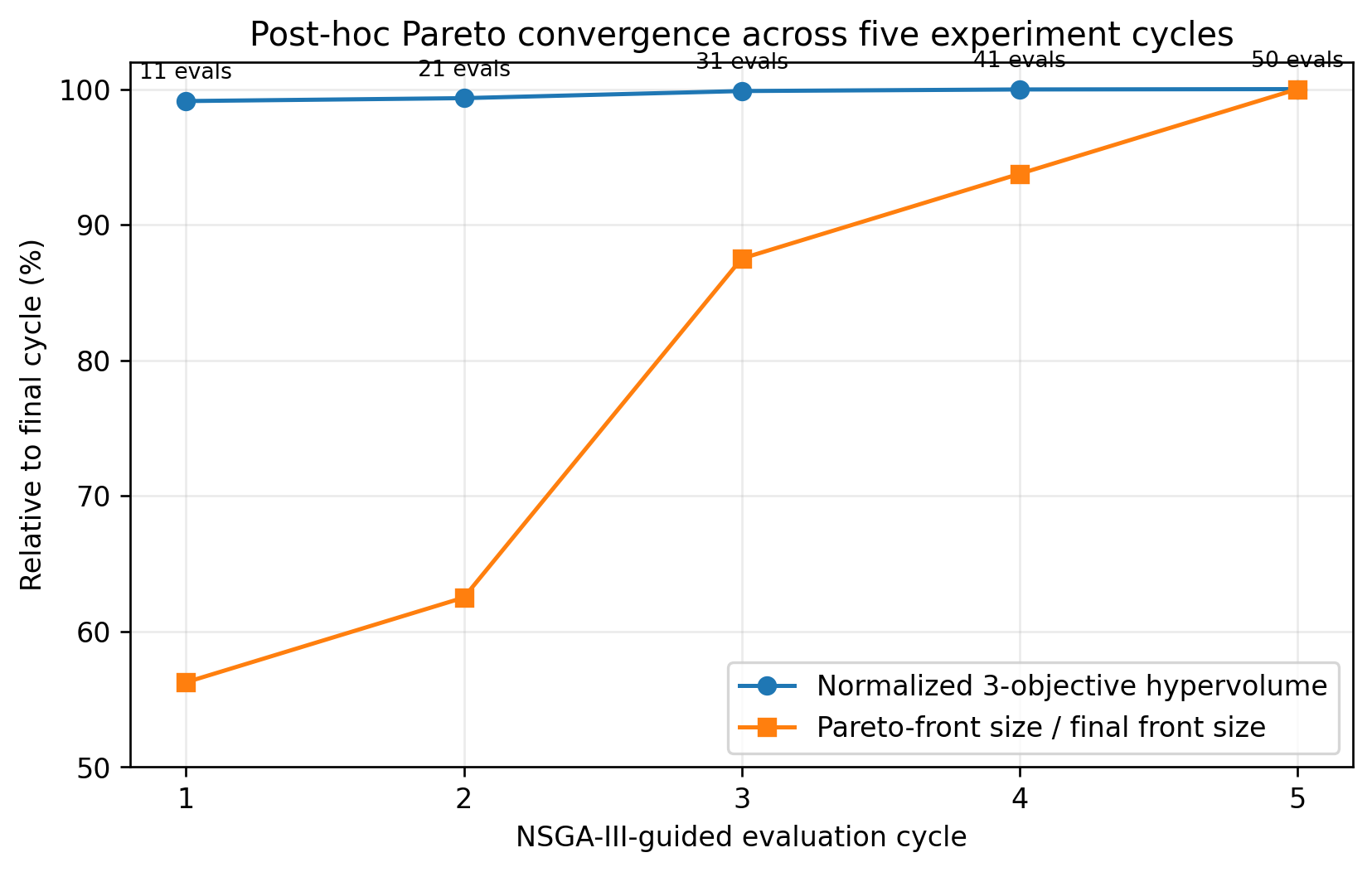}
\caption{Post-hoc search convergence.  Hypervolume is normalized to the final cycle; Pareto-front size is shown as a fraction of the final 16-point front.}
\label{fig:pareto_convergence}
\end{figure}

\subsection{INT8 and INT4 deployment implementation}
\label{subsec:quant_impl}

\par\medskip
\noindent\textbf{Training-time QAT versus deployed precision.}
\par\smallskip
The $x_Q$ training modes expose the model to fake-quantized values during optimization, but fake quantization alone does not create packed INT8 or 4-bit weights.  Deployment conversion is therefore treated separately from QAT.  This distinction is essential for both the resident-memory and RTF results.

\par\medskip
\noindent\textbf{INT8.}
\par\smallskip
Two INT8 deployment paths were used for different questions.  The historical resident-memory artifacts use genuine packed INT8 conversion for supported Linear layers, so their static memory is a real converted-artifact measurement.  For optimized GPU runtime, the trained Whisper-Small LoRA-$r64$ checkpoint is loaded, the LoRA update is merged, and the resulting Hugging Face model is converted once to CTranslate2.  The same serialized CTranslate2 model can then be loaded as \texttt{float16} or \texttt{int8\_float16}; no retraining is required.  For the $x_V=2$ checkpoint the CTranslate2 encoder is executed first, its GPU-resident hidden states are subsampled by two, and the result is passed directly to the CTranslate2 decoder.  The measured timing scope includes encoder execution, GPU subsampling, and generation, while feature extraction and the host-to-device feature copy are excluded.  The optimized INT8 test reported an RTF of approximately 0.0262.  Since the paired CTranslate2 FP16 summary is not included in the available experiment record here, we report 0.0262 as an optimized-backend absolute RTF and do not infer a pure INT8 speedup by dividing it by the PyTorch FP16 number.

\par\medskip
\noindent\textbf{INT4/NF4.}
\par\smallskip
The 4-bit deployment path is bitsandbytes NF4.  The actual fine-tuned weights are first merged into a flat model, serialized, and then reloaded with \texttt{load\_in\_4bit = True},  \texttt{bnb\_4bit\_quant\_type = ``nf4''}, double quantization enabled, and \texttt{bnb\_4bit\_compute dtype = float16}.  NF4 stores each quantized weight as one of 16 non-uniform code values chosen for normally distributed neural-network weights.  Two 4-bit codes are packed per physical byte.  However, the runtime dequantizes into the FP16 computation path; consequently NF4 is a genuine 4-bit \emph{weight-storage} implementation but not native W4A4 INT4 arithmetic.  The $q_Q=0.125$ EffFLOPs factor therefore represents an idealized optimized-INT4 compute regime, not the literal bitsandbytes execution path.

A controlled same-checkpoint experiment makes this backend dependence explicit.  The same dense Small LoRA-$r64$ checkpoint was deployed as FP32, FP16, bitsandbytes INT8, and bitsandbytes NF4.  Each condition used the first 500 LibriSpeech test-other examples, three fresh-process repeats, five warm-up calls, beam size 16, KV caching, \texttt{max\_new\_tokens=256}, \texttt{no\_repeat\_ngram\_size=3}, and repetition penalty 1.3 on the same RTX A6000.  Timing used synchronized CUDA events and wall clock around model generation; feature extraction was excluded.

\begin{table*}[t]
\centering
\caption{Controlled precision/backend ablation on the same trained Small LoRA-$r64$ checkpoint.  Peak GPU allocation is a runtime diagnostic and is distinct from the paper's resident-weight-memory objective.}
\label{tab:rtf_quant}
\scriptsize
\setlength{\tabcolsep}{4pt}
\begin{tabular}{@{}lllrrrr@{}}
\toprule
Deployment & Backend & WER (\%) & Wall RTF & p50 (ms) & p95 (ms) & Peak alloc. (GB)\\
\midrule
FP32 & PyTorch dense, TF32 off & 6.7657 & $0.053397\pm0.000162$ & 305.869 & 783.306 & 2.0904\\
FP16 & PyTorch dense & 6.7336 & $0.045363\pm0.000134$ & 252.392 & 703.803 & 1.0720\\
INT8 & bitsandbytes LLM.int8 mixed path & 7.1398 & $0.180516\pm0.000177$ & 1034.541 & 2828.696 & 0.9287\\
NF4 & bitsandbytes 4-bit weights / FP16 compute & 7.1932 & $0.061355\pm0.000055$ & 352.776 & 948.312 & 0.7599\\
INT8 & CTranslate2 optimized path & -- & $\approx0.0262$ & -- & -- & --\\
\bottomrule
\end{tabular}
\end{table*}

The controlled PyTorch/bitsandbytes experiment shows that lower nominal precision does not guarantee lower latency: FP16 is about $1.18\times$ faster than FP32, while the tested bitsandbytes INT8 and NF4 paths are slower than FP32 despite reducing GPU allocation.  This is not a contradiction in the EffFLOPs definition; it demonstrates that EffFLOPs is an analytical search surrogate whose hardware realization depends on kernel and serving implementation.  The much lower CTranslate2 INT8 RTF illustrates why backend choice must be separated from the nominal precision label.

\begin{figure}[t]
\centering
\includegraphics[width=\columnwidth]{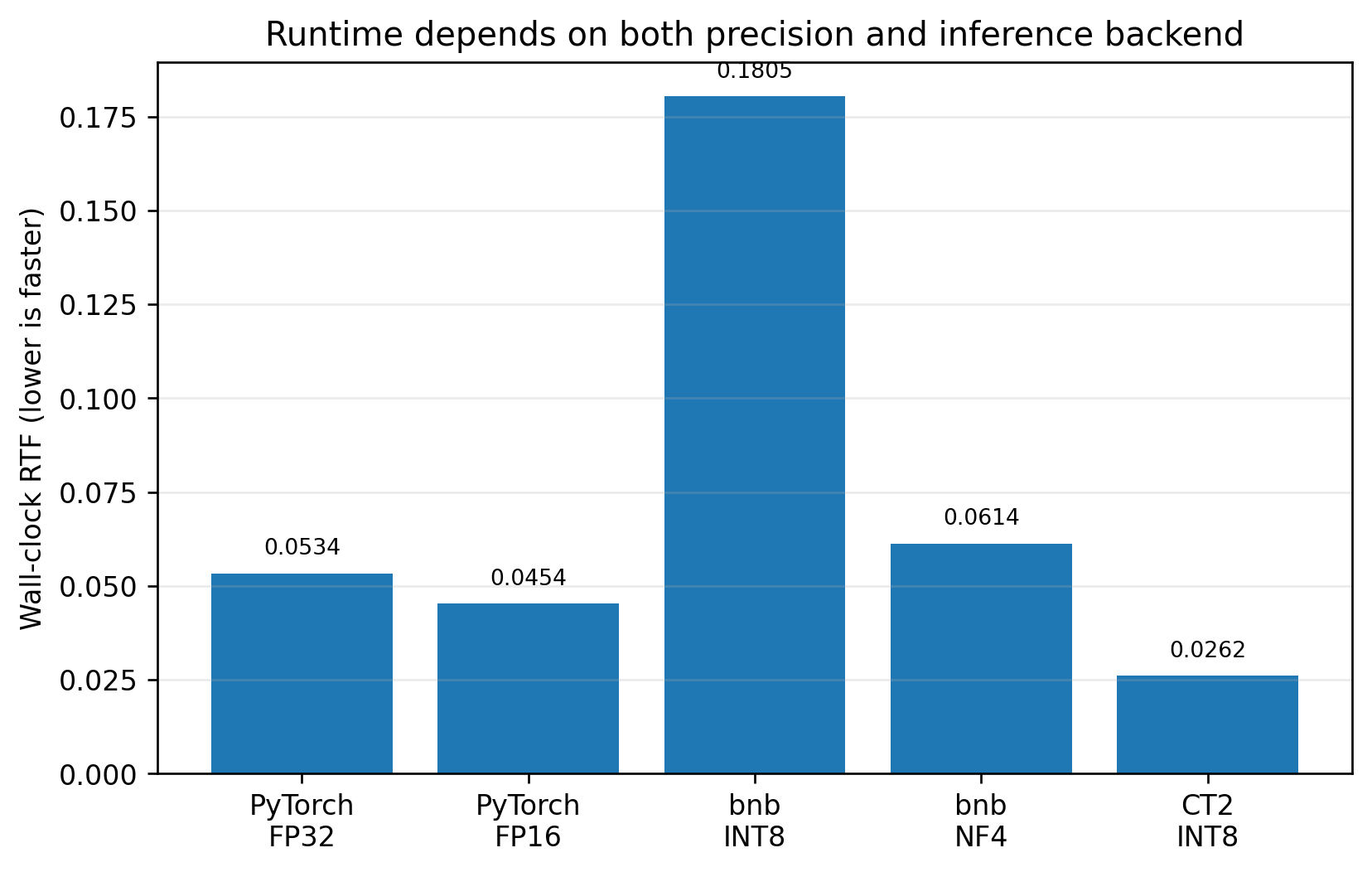}
\caption{Measured RTF under several precision/backend combinations.  The CTranslate2 INT8 bar is intentionally labeled by backend because it is not a same-backend continuation of the PyTorch/bitsandbytes bars.}
\label{fig:precision_rtf}
\end{figure}

\subsection{N:M structured sparsity implementation}
\label{subsec:sparsity_impl}

The implementation uses the conventional interpretation that $N$ in $N$:$M$ denotes the number of \emph{retained non-zero} weights in each group of $M$.  Thus 2:4 retains two and prunes two weights per group (50\% sparsity), whereas 1:4 retains one and prunes three (75\% sparsity).  This wording matters: describing ``1:4'' as pruning one of four would be inconsistent with the code and with the actual masks.

Pruning is applied to eligible base-model \texttt{nn.Linear} matrices only.  LoRA A/B matrices are excluded, the tied vocabulary output projection is excluded, and matrices whose dimensions are incompatible with four-element grouping are skipped.  Each eligible matrix is reshaped into row-wise groups of four; weights are ranked by absolute magnitude and the smallest $M-N$ elements are zeroed.  For LoRA recovery the sparse base remains fixed while the dense adapter is trained.  The 2:4 path uses a one-shot mask, while the 1:4 recovery path can recompute the survivor mask every 50 steps during the initial 500-step selection window before freezing it.

For standard Whisper blocks, each encoder layer contributes six eligible matrices (Q/K/V/out and two FFN matrices), and each decoder layer contributes ten (four self-attention, four cross-attention, and two FFN matrices).  Hence
\begin{equation}
N_{\mathrm{layers}}=6N_E+10N_D,
\end{equation}
and, because $d_{ff}=4d$ for these checkpoints, the eligible scalar-weight count is
\begin{equation}
N_{\mathrm{eligible}}=(12N_E+16N_D)d^2.
\end{equation}
Table~\ref{tab:prune_counts} gives the resulting counts.  These are individual weight elements, not entire neurons or Transformer blocks.

\begin{table*}[t]
\centering
\caption{Linear weights affected by N:M pruning.  Whole-model percentages use the nominal released parameter counts and therefore are approximate.}
\label{tab:prune_counts}
\scriptsize
\begin{tabular}{@{}lrrrrrr@{}}
\toprule
Model & Eligible Linear layers & Eligible weights & 2:4 zeroed & 1:4 zeroed & 2:4 / all params & 1:4 / all params\\
\midrule
Tiny & 64 & 16.52M & 8.26M & 12.39M & 21.2\% & 31.8\%\\
Small & 192 & 198.18M & 99.09M & 148.64M & 40.6\% & 60.9\%\\
Medium & 384 & 704.64M & 352.32M & 528.48M & 45.8\% & 68.7\%\\
Large-v3 & 512 & 1468.01M & 734.00M & 1101.00M & 47.7\% & 71.5\%\\
Distil-Small & 112 & 122.68M & 61.34M & 92.01M & 37.0\% & 55.4\%\\
Distil-Medium & 164 & 335.54M & 167.77M & 251.66M & 42.6\% & 63.9\%\\
Distil-Large & 212 & 681.57M & 340.79M & 511.18M & 45.1\% & 67.6\%\\
\bottomrule
\end{tabular}
\end{table*}

A second controlled runtime experiment used the same recovered 2:4 Small LoRA-$r64$ checkpoint in two forms: (i) a zero-masked FP16 tensor still executed through dense GEMM and (ii) a PyTorch semi-structured 2:4 base with the dense LoRA branch left unmerged.  Exactly 192 eligible Small-model Linear layers were converted in the latter condition.  Under this batch-1 autoregressive workload the semi-structured PyTorch path was slower, not faster: RTF increased from $0.058633\pm0.000471$ to $0.282771\pm0.006487$.  Peak allocation decreased from 1.0913 to 0.9010 GB.  The result should not be generalized to all 2:4 hardware implementations; rather, it is direct evidence that the analytical 0.5 density factor is not by itself a latency model for this particular sparse kernel/workload combination.  No native 1:4 runtime kernel was implemented.

\begin{figure}[t]
\centering
\includegraphics[width=\columnwidth]{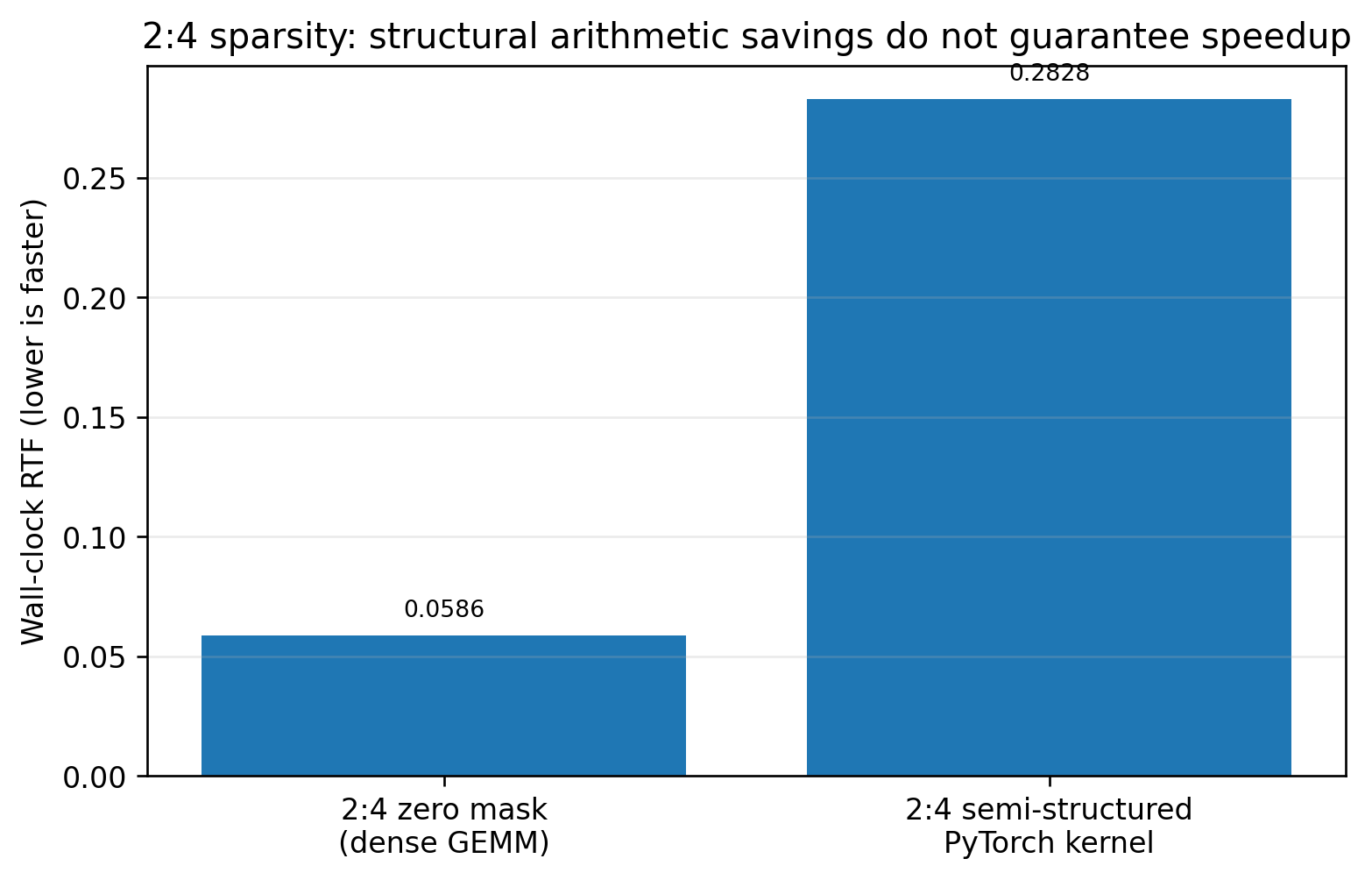}
\caption{Controlled 2:4 runtime ablation.  The same recovered sparse checkpoint is executed either as a dense zero-masked tensor or through PyTorch's semi-structured 2:4 path.}
\label{fig:sparsity_rtf}
\end{figure}

\subsection{Extended 1:4 recovery ablation}
\label{subsec:onefour_recovery}

The 1:4 pattern is the most destructive structural setting because three of every four eligible Linear weights are removed.  Representative failures span model scale, adaptation mode, and precision: Large-v3 Distil/full/FP32 reaches 97.83\% WER; Small Distil/$r=16$/INT4 reaches 139.05\%; Small Distil/$r=8$/INT8 reaches 183.26\%; and Tiny/$r=64$/FP32 reaches 102.78\%.  To test whether the standard 500-step recovery window was simply too short, a full-parameter Medium/FP32 configuration was extended to 4000 recovery steps.  The teacher-forced training loss fell monotonically from 4.915 at step 800 to 4.178 at step 4000, while WER rose from 94.4\% to 104.0\%.

\begin{table}[t]
\centering
\caption{Extended 1:4 full-parameter Medium recovery.}
\label{tab:onefour_recovery}
\scriptsize
\begin{tabular}{@{}rrr@{}}
\toprule
Step & Training loss & WER (\%)\\
\midrule
800  & 4.915 & 94.4\\
1600 & 4.365 & 99.3\\
2400 & 4.240 & 103.7\\
3200 & 4.192 & 103.9\\
4000 & 4.178 & 104.0\\
\bottomrule
\end{tabular}
\end{table}

\begin{figure}[t]
\centering
\includegraphics[width=\columnwidth]{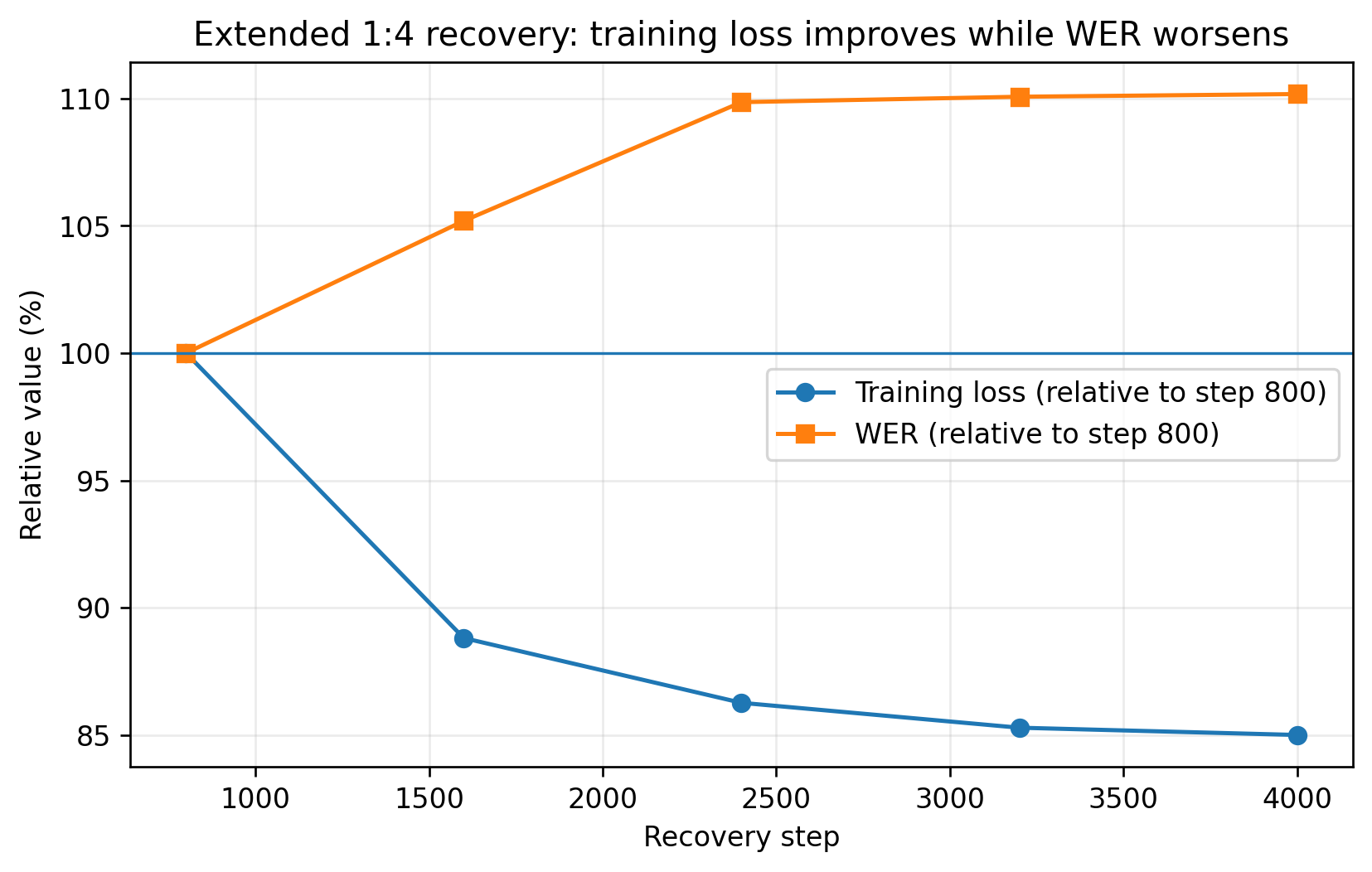}
\caption{The 1:4 extended-recovery diagnostic normalized to the first reported point.  Optimization continues to reduce the training loss while free-running WER deteriorates, so merely extending the recovery budget did not resolve the failure in this experiment.}
\label{fig:onefour_recovery}
\end{figure}

This result supports the paper's practical conclusion only within the tested recovery budget.  It does not prove that 1:4 is fundamentally unrecoverable at substantially larger training scale.  In addition, some early sweep evaluations exhibited generation-configuration/repetition sensitivity; the 4000-step full-parameter trajectory is therefore the cleaner ablation for arguing that additional recovery optimization alone was insufficient.

\subsection{Resident weight memory}
\label{subsec:resident_memory}

The memory objective is \emph{resident weight memory}, not peak runtime GPU memory.  It is computed from the physical storage of parameters and buffers in the actual inference object,
\begin{equation}
M_{\mathrm{resident}}=\frac{\sum_i n_i b_i+\sum_j n_j b_j}{1024^2}\ \mathrm{MB},
\end{equation}
where $n$ is the stored element count and $b$ is the physical bytes per stored element.  Dense FP32 and FP16/BF16 tensors contribute four and two bytes per element respectively; packed INT8 tensors contribute one byte per stored element.  For bitsandbytes NF4, \texttt{Params4bit} is already packed into a \texttt{uint8} container that stores two 4-bit codes per byte, so its reported packed element count is already compressed.  Applying an additional factor of $0.5$ to that packed \texttt{uint8} representation would double-count the compression.

This metric intentionally excludes activations, KV cache, beam-search state, temporary workspaces, and allocator reservation.  Those quantities are workload dependent and are separately visible in the controlled RTF experiment's peak-allocation column.  Zero-masking a dense 2:4 or 1:4 tensor also does not automatically reduce resident storage: a zero is still an FP16/FP32 element unless the weight is actually stored in a compressed sparse representation.

The quantized-memory behavior can be understood with a two-pool model.  Let $L$ denote Linear-layer parameters touched by quantization and $R$ all remaining parameters.  Ignoring small metadata terms,
\begin{align}
M_{\mathrm{FP16}} &\approx 2L+2R,\\
M_{\mathrm{NF4}}  &\approx 0.5L+2R,\\
M_{\mathrm{INT8,hist}} &\approx 1L+4R,
\end{align}
where the last expression describes the project's historical dynamic-INT8 artifact path in which non-Linear tensors remained FP32.  This explains why real converted artifacts do not achieve the naive whole-model $4\times$ or $2\times$ storage reductions.  In the main results, Medium decreases from 1492.9 MB in dense FP16 to 897.9 MB under the real INT8 artifact path ($0.60\times$), while Small decreases from 474.6 MB to 177.6 MB under the real NF4 deployment ($0.37\times$).  These are physical artifact measurements, not precision multipliers.

\subsection{What the runtime ablations imply for EffFLOPs}
\label{subsec:effflops_limits}

The runtime experiments deliberately expose the boundary of the EffFLOPs abstraction.  The analytical factors remain useful for ordering a very large configuration space without benchmarking every candidate on every deployment backend, but they are not a hardware-independent latency predictor.  On the tested PyTorch/bitsandbytes stack, the INT8 and NF4 paths do not realize the nominal $0.25$/$0.125$ timing ratios, and the tested PyTorch semi-structured 2:4 kernel is slower than dense-zero execution for this batch-1 autoregressive workload.  Conversely, the optimized CTranslate2 INT8 path reaches an absolute RTF around 0.0262, demonstrating that implementation quality can move runtime much closer to the intended low-precision deployment regime.  Native optimized INT4 arithmetic and a production-quality 1:4 sparse kernel remain outside the implemented scope.

The correct interpretation of the three-objective search is therefore
\begin{center}
\textbf{Pareto search = WER + analytical EffFLOPs + measured resident memory.}
\end{center}
RTF serves as a separate implementation-dependent validation layer rather than being silently equated with EffFLOPs.

\end{document}